\documentclass{amsart}
\usepackage[dvips]{graphicx}
\usepackage{epsfig,enumerate,mathrsfs}
\usepackage[notref,notcite]{showkeys}
\usepackage{color}

\numberwithin{equation}{section}

\theoremstyle{definition}

\newcommand{\R}{\mathbb R}
\newcommand{\Z}{\mathbb Z}

\def\N{\mathbb{N}}
\def\C{\mathbb{C}}

\def\KD3{\mathrm{KD}^3}

\def\beq{\begin{eqnarray}}
\def\eeq{\end{eqnarray}}
\newcommand{\nn}{\nonumber}

\newcommand{\snmp}{\sum_{n=-\infty}^\infty}
\newcommand{\slnp}{\sum_{\ell =0}^\infty}
\newcommand{\sln}{\sum_{n=-\infty}^\infty \sum_{\ell =0}^\infty}
\newcommand{\pnnu}{\left( 1+\frac{p_n^2}{\nu^2}\right)}
\newcommand{\mnup}{\frac{m^2}{\nu^2+p_n^2}}
\newcommand{\pmnu}{\frac{p_n^2+m^2}{\nu^2}}

\begin{document}

\title[Functional determinants in higher dimensions]{Functional determinants in higher dimensions using contour integrals}

\author{Klaus Kirsten}
\address{Department of Mathematics\\ Baylor University\\
         Waco\\ TX 76798\\ U.S.A. }
\email{Klaus$\_$Kirsten@baylor.edu}

\begin{abstract}
In this contribution we first summarize how contour integration methods can be used to derive closed formulae for functional determinants of ordinary differential operators. We then generalize our considerations to partial differential operators. Examples are used to show that also in higher dimensions closed answers can be obtained as long as the eigenvalues of the differential operators are determined by transcendental equations. Examples considered comprise of the finite temperature Casimir effect on a ball and the functional determinant of the Laplacian on a two-dimensional torus.
\end{abstract}

\maketitle

\section{Introduction}
Functional determinants of second order differential operators are of great importance in many different fields. In physics, functional determinants provide the so-called one-loop approximation to quantum field theories in the path integral formulation \cite{feyn65b,schu81b}. In mathematics it describes the analytical torsion of a manifold \cite{ray71-7-145}.

Although there are various ways to evaluate functional determinants, the zeta function scheme seems to be the most elegant technique to use \cite{byts96-266-1,eliz95b,eliz94b,kirs01b}. This is the method introduced by Ray and Singer to define analytical torsion \cite{ray71-7-145}. In physics its origin goes back to ambiguities in dimensional regularization when applied to quantum field theory in curved spacetime \cite{dowk76-13-3224,hawk77-55-133}.

For many second order ordinary differential operators surprisingly simple answers can be given. The determinants for these situations have been related to boundary values of solutions of the operators, see, e.g., \cite{burg91-138-1,cole85b,drey78-45-15,form87-88-447,form92-147-485,gelf60-1-48,klei06b,lesc98-194-139,lesc98-193-643}. Recently, these results have been rederived with a simple and accessible method which uses contour integration techniques \cite{kirs08-76-60,kirs03-308-502,kirs04-37-4649}. The main advantage of this approach is that it can be easily applied to general kinds of boundary conditions \cite{kirs04-37-4649} and also to cases where the operator has zero modes \cite{kirs03-308-502,kirs04-37-4649}; see also \cite{klei98-245-345,klei99-40-6044,mcka95-28-6931}. Equally important, for some higher dimensional situations the task of finding functional determinants remains feasible. Once again closed answers can be found but compared to one dimension technicalities are significantly more involved \cite{dunn06-39-11915,dunn09-42-075402}. It is the aim of this article to choose specific higher dimensional examples where technical problems remain somewhat confined. The intention is to illustrate that also for higher dimensional situations closed answers can be obtained which are easily evaluated numerically.

The outline of this paper is as follows. In Section 2 the essential ideas are presented for ordinary differential operators. In Section 3 and 4 examples of functional determinants for partial differential operators are considered. The determinant in Section 3 describes the finite temperature Casimir effect of a massive scalar field in the presence of a spherical shell \cite{defr94-50-2908,defr97-55-2477}. The calculation in Section 4 describes determinants for strings on world-sheets that are tori \cite{polc98b,will03b} and it gives an alternative derivation of known results. Section 5 summarizes the main results.

\section{Contour integral formulation of zeta functions}\label{sec2}
In this section we review the basic ideas that lead to a suitable contour integral representation of zeta functions associated with ordinary differential operators. This will form the basis of the considerations for partial differential operators to follow later.

We consider the simple class of differential operators $$P:= -\frac{d^2}{dx^2} + V(x)$$ on the interval $I=[0,1]$, where $V(x)$ is a smooth potential. For simplicity we consider Dirichlet boundary conditions. From spectral theory \cite{levi75b} it is known that there is a spectral resolution $\{\phi_n , \lambda_n\}_{n=1}^\infty$ satisfying $$P\phi_n (x) = \lambda_n \phi _n (x), \quad \quad \phi_n (0) = \phi _n (1) =0.$$ The spectral zeta function associated with this problem is then defined by \beq \zeta _P (s) = \sum_{n=1}^\infty \lambda_n^{-s}, \label{res01}\eeq
where by Weyl's theorem about the asymptotic behavior of eigenvalues \cite{weyl12-71-441} this series is known to converge for $\Re s > 1/2$.

If the potential is not a very simple one, eigenfunctions and eigenvalues will not be known explicitly. So how can the zeta function in equation (\ref{res01}), and in particular the determinant of $P$ defined via $$\det P = e^{-\zeta_P ' (0)},$$ be analyzed? From complex analysis it is known that series can often be evaluated with the help of the argument principle or Cauchy's residue theorem by rewriting them as contour integrals. In the given context this can be achieved as follows. Let $\lambda \in \C$ be an arbitrary complex number. From the theory of ordinary differential equations it is known that the initial value problem \beq (P-\lambda )u_\lambda (x) =0, \quad \quad u_\lambda (0) =0, \quad u_\lambda ' (0) =1,\label{res01a}\eeq has a unique solution. The connection with the boundary value problem is made by observing that the eigenvalues $\lambda_n$ follow as solutions to the equation \beq u_\lambda (1) =0;\label{res02}\eeq
note that $u_\lambda (1)$ is an analytic function of $\lambda$.

With the help of the argument principle, equation (\ref{res02}) can be used to write the zeta function, equation (\ref{res01}), as \beq \zeta_P (s) = \frac 1 {2\pi i} \int\limits_\gamma d\lambda \,\, \lambda^{-s} \frac d {d\lambda} \ln u_\lambda (1).\label{res03}\eeq
Here, $\gamma$ is a counterclockwise contour and encloses all eigenvalues which we assume to be positive, see Figure 1. The pertinent remarks when finitely many eigenvalues are non-positive are given in \cite{kirs04-37-4649}.

\begin{figure}[ht]
\setlength{\unitlength}{1cm}

\begin{center}
\begin{picture}(10,6.5)
\thicklines \put(0,3){\vector(1,0){10}}
\put(5.0,0){\vector(0,1){6}}
\multiput(5.6,3)(.4,0){10}{\circle*{.15}}
\qbezier(5,3.01)(4,3.01)(0,3.01) \qbezier(5,2.99)(4,2.99)(0,2.99)
\qbezier(5,2.97)(4,2.97)(0,2.97) \qbezier(5,3.03)(4,3.03)(0,3.03)
\put(8.0,5.5){{\bf\huge $\lambda$-plane}}
\thicklines\put(9.7,3.0){\oval(9.0,1.0)[l]}
\put(8.0,3.5){\vector(-1,0){.8}}

\end{picture}
\caption{Contour $\gamma$ used in equation (\ref{res03}).}
\end{center}
\end{figure}
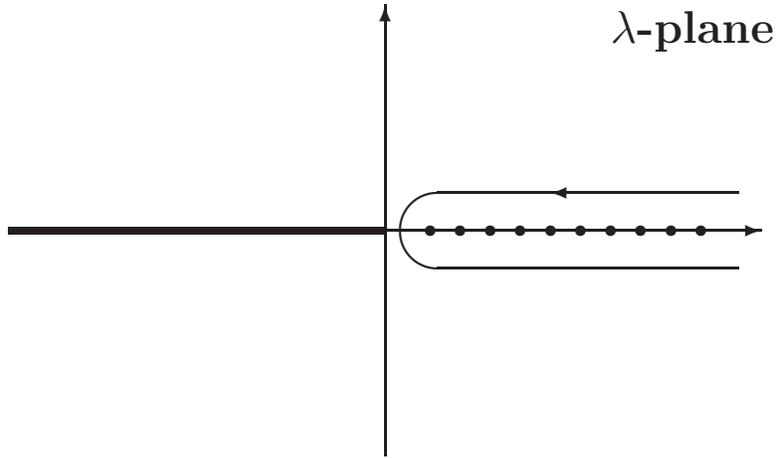

The asymptotic behavior of $u_\lambda (1)$ as $|\lambda |\to \infty$, namely $$u_\lambda (1) \sim \frac{\sin \sqrt \lambda}{\sqrt \lambda}, $$ implies that this representation is valid for $\Re s > 1/2$. In order to find the determinant of $P$ we need to construct the analytical continuation of equation (\ref{res03}) to a neighborhood about $s=0$. This is best done by deforming the contour to enclose the branch cut along the negative real axis and then shrinking it to the negative real axis, see Figure 2.

\begin{figure}[ht]
\setlength{\unitlength}{1cm}

\begin{center}
\begin{picture}(10,6.5)
\thicklines

\put(0,3){\vector(1,0){10}} \put(5.0,0){\vector(0,1){6}}
\put(0.0,3.0){\oval(10.5,1.0)[r]}  \put(2.0,3.5){\vector(1,0){.4}}
\multiput(5.6,3)(.4,0){10}{\circle*{.1}}
\qbezier(5,3.01)(4,3.01)(0,3.01) \qbezier(5,2.99)(4,2.99)(0,2.99)
\qbezier(5,2.97)(4,2.97)(0,2.97) \qbezier(5,3.03)(4,3.03)(0,3.03)
\put(8.0,5.5){{\bf\huge $\lambda$-plane}}


\end{picture}
\caption{Contour $\gamma$ used in equation (\ref{res03}) after deformation.}
\end{center}
\end{figure}
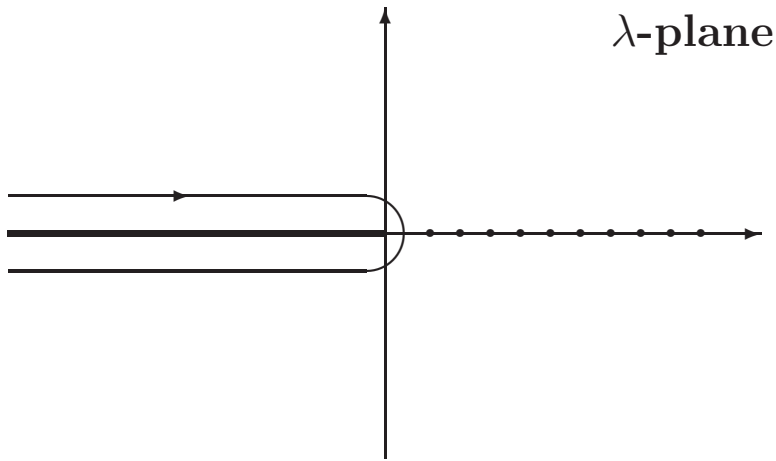

The outcome is \beq \zeta_P (s) = \frac{\sin \pi s} \pi \int\limits_0^\infty d\lambda \,\, \lambda^{-s} \frac d {d\lambda} \ln u_{-\lambda } (1) . \label{res04}\eeq
To see where this representation is well defined notice that for $\lambda \to \infty$ the behavior follows from \cite{levi75b} $$u_{-\lambda } (1) \sim \frac{\sin (i\sqrt \lambda )}{i\sqrt \lambda} = \frac{e^{\sqrt \lambda}}{2\sqrt \lambda} \left( 1-e^{-2\lambda}\right). $$ The integrand, to leading order in $\lambda$, therefore behaves like $\lambda^{-s-1/2}$ and convergence at infinity is established for $\Re s > 1/2$. As $\lambda \to 0$ the behavior $\lambda^{-s}$ follows. Therefore, in summary, (\ref{res04}) is well defined for $1/2<\Re s <1$. In order to shift the range of convergence to the left we add and subtract the leading $\lambda\to\infty$ asymptotic behavior of $u_{-\lambda} (1) $. The whole point of this procedure will be to obtain one piece that at $s=0$ is finite, and another piece for which the analytical continuation can be easily constructed.

Given we want to improve the $\lambda \to \infty$ behavior without worsening the $\lambda \to 0$ behavior, we split the integration range. In detail we write \beq \zeta_P (s) = \zeta_{P,f} (s) + \zeta_{P,as} (s) , \label{res05}\eeq
where \beq \zeta_{P,f} (s) &=& \frac{\sin \pi s} \pi \int\limits_0^1 d\lambda \,\, \lambda ^{-s} \frac d {d\lambda} \ln u_{-\lambda} (1)\nn\\
&+&\frac{\sin \pi s} \pi \int\limits_1^\infty d\lambda \,\, \lambda ^{-s} \frac d {d\lambda} \ln \left( u_{-\lambda} (1) 2 \sqrt \lambda e^{-\sqrt \lambda}  \right), \label{res06} \\
\zeta_{P,as} (s) &=& \frac{\sin \pi s} \pi \int\limits_1^\infty d\lambda \,\, \lambda ^{-s} \frac d {d\lambda} \ln \left( \frac{e^{\sqrt \lambda}}{2\sqrt \lambda}\right).\label{res07}\eeq
By construction, $\zeta_{P,f} (s)$ is analytic about $s=0$ and its derivative at $s=0$ is trivially obtained, \beq \zeta_{P,f} ' (0) = \ln u_{-1} (1) - \ln u_0 (1) - \ln \left( u_{-1} (1) 2 e^{-1} \right) = - \ln \left( \frac{ 2u_0 (1)} e \right) .\label{res08} \eeq
Although the representation (\ref{res07}) is only defined for $\Re s > 1/2$, the analytic continuation to a meromorphic function on the complex plane is found using $$\int\limits_1^\infty d\lambda \,\, \lambda^{-\alpha} = \frac 1 {\alpha -1}\quad \quad \mbox{for}\quad \quad \Re \alpha >1.$$ This shows \beq \zeta_{P,as} (s) = \frac{\sin \pi s} {2\pi} \left( \frac 1 {s-\frac 1 2 } - \frac 1 s \right) ,\nn\eeq and furthermore \beq \zeta_{P,as} ' (0) = -1.\nn\eeq Adding up, the final answer reads \beq \zeta_P ' (0) = - \ln (2u_0 (1)).\label{res08a}\eeq
For the numerical evaluation of the determinant, not even one eigenvalue is needed. The only relevant information is the boundary value of the unique solution to the initial value problem $$\left( - \frac{d^2}{dx^2} + V(x) \right) u_0 (x) =0, \quad \quad u_0 (0) =0, \quad \quad u_0 ' (0) =1.$$

General boundary conditions can be dealt with as easily. The best formulation results by rewriting the second order differential equation as a first order system in the usual way. Namely, we define $v_\lambda (x) = du_\lambda (x) / dx$ such that the differential equation (\ref{res01a}) turns into \beq \frac d {dx} {u_\lambda (x) \choose v_\lambda (x) } = \left(\begin{array}{cc} 0 & 1\\ V(x) -\lambda & 0 \end{array}\right) {u_\lambda (x) \choose v_\lambda (x) }.\label{res09}\eeq Linear boundary conditions are given in the form \beq M {u_\lambda (0) \choose v_\lambda (0) } + N {u_\lambda (1) \choose v_\lambda (1) } = {0 \choose 0}, \label{res010}\eeq
where $M$ and $N$ are $2\times 2$ matrices whose entries characterize the nature of the boundary conditions. For example, the previously described Dirichlet boundary conditions are obtained by choosing $$ M = \left( \begin{array}{cc} 1 & 0 \\ 0 & 0 \end{array} \right), \quad \quad
N = \left( \begin{array}{cc} 0 & 0 \\ 1 & 0 \end{array} \right) . $$
In order to find an implicit equation for the eigenvalues like equation (\ref{res02}) we use the fundamental matrix of (\ref{res09}). Let $u_\lambda^{(1)} (x)$ and $u_\lambda^{(2)} (x)$ be linearly independent solutions of (\ref{res09}). Suitably normalized, these define the fundamental matrix
\beq H_\lambda (x) = \left( \begin{array}{cc} u_\lambda^{(1)} (x) & u_\lambda ^{(2)} (x) \\
v_\lambda^{(1)} (x) & v_\lambda ^{(2)} (x) \end{array} \right), \quad \quad H_\lambda (0) = \mbox{Id}_{2\times 2}.\nn\eeq
The solution of (\ref{res09}) with initial value $(u_\lambda (0), v_\lambda (0))$ is then obtained as $${u_\lambda (x) \choose v_\lambda (x) } = H_\lambda (x) {u_\lambda (0) \choose v_\lambda (0) } .$$ The boundary conditions (\ref{res010}) can therefore be rewritten as \beq (M+NH_\lambda (1) ) {u_\lambda (0) \choose v_\lambda (0) } = {0\choose 0} .\label{res011}\eeq
This shows that the condition for eigenvalues to exist is $$\det (M+NH_\lambda (1) ) =0,$$ which replaces (\ref{res02}) in case of general boundary conditions. The zeta function associated with the boundary condition (\ref{res010}) therefore takes the form $$\zeta_P (s) = \frac 1 {2\pi i} \int\limits_\gamma d\lambda \,\, \lambda^{-s} \frac d {d\lambda} \ln \det (M+NH_\lambda (1)) $$ and the analysis proceeds from here depending on $M$ and $N$.  If $P$ represents a system of operators one can proceed along the same lines. Note, that we have replaced the task of evaluating the determinant of a differential operator by one of computing the determinant of a finite matrix.

The procedure just outlined is by no means confined to be applied
to ordinary differential operators only. In fact, the zeta function associated with many boundary value
problems allowing for a separation of variables can be analyzed
using this contour integral technique. In more detail, starting off with some coordinate system, see,
e.g., \cite{moon61b}, eigenvalues are often determined by
$$F_j (\lambda_{j,n}) =0 , $$ where $j$ is a suitable quantum
number depending on the coordinate system considered and $F_j$ is
a given special function depending on the coordinate system; e.g.
for ellipsoidal coordinate systems the relevant special function
is the Mathieu function. Continuing along the lines described above,
denoting by $d_j$ an appropriate degeneracy that might be present,
we write somewhat symbolically \beq\zeta_P (s) = \sum_j d_j \frac 1 {2\pi i}
\int\limits_\gamma d\lambda  \lambda^{-s} \frac \partial
{\partial \lambda} \ln F_j (\lambda ) , \label{ns5}\eeq the task being to
construct the analytical continuation of this object to
$s=0$. The details of the procedure will depend very much
on the properties of the special function $F_j$ that enters. For
example, on balls Bessel functions are relevant \cite{bord96-37-895,bord96-179-215,bord96-182-371},
the spherical suspension \cite{barv92-219-201}, or
sphere-disc configurations \cite{gilk01-601-125,kirs02-104-119},
involve Legendre functions, ellipsoidal boundaries involve Mathieu
functions etc. For many examples relevant properties of $F_j (\lambda )$ are
not available in the literature and need to be derived using
techniques of asymptotic analysis
\cite{levi75b,olve54-247-328,olve74b}. For quite common coordinate
systems like the polar coordinates this is not necessary. When the asymptotics is
known, the relevant integrals resulting in (\ref{ns5}) need to be
evaluated and closed expressions representing the determinant of partial differential operators are found. Although
the remaining sums in general cannot be explicitly performed, the results obtained are very suitable for numerical evaluation.

\section{Finite temperature Casimir energy on the ball}
Let us now apply the above remarks about higher dimensions using the general formalism described in \cite{dunn09-42-075402}. As  a concrete example we consider the finite temperature theory of a massive scalar field on the three dimensional ball. Using the zeta function scheme we have to consider the eigenvalue problem \beq P \phi_\lambda (\tau , \vec x ) := \left( - \frac{d^2} {d\tau ^2} - \Delta + m^2 \right) \phi_\lambda (\tau , \vec x ) = \lambda ^2 \phi_\lambda (\tau , \vec x) , \label{res1}\eeq
where $\tau$ is the imaginary time and $\vec x \in B^3 := \{\vec x \in \R^3| |\vec x| \leq 1\}.$ We have written $\lambda^2$ for the eigenvalues to avoid the occurrence of square roots in arguments of Bessel functions later on.

For finite temperature theory we impose periodic boundary conditions in the imaginary time, \beq \phi_\lambda (\tau, \vec x) = \phi_\lambda (\tau + \beta , \vec x) , \nn\eeq
where $\beta$ is the inverse temperature, and for simplicity we choose Dirichlet boundary conditions on the boundary of the ball, \beq \left.\phi_\lambda (\tau, \vec x) \right|_{| \vec x| =1} =0. \nn\eeq
The zeta function associate with this boundary value problem is then \beq \zeta_P (s) = \sum_\lambda \lambda^{-2s} , \label{res2}\eeq
and the energy of the system is defined by \beq E := - \frac 1 2 \frac{\partial} {\partial \beta} \zeta_{P/\mu^2} ' (0) , \label{res3}\eeq
where $\mu$ is an arbitrary parameter with dimension of a mass introduced in order to get the correct dimension for the energy.
For a full discussion of its relevance in the renormalization process in this model at zero temperature see \cite{bord97-56-4896}. That discussion remains completely unchanged at finite temperature and we will put $\mu =1$ henceforth.

Given the radial symmetry of the problem we separate variables in polar coordinates according to \beq \phi_\lambda (\tau, r, \theta ,\varphi ) = \frac 1 {\sqrt r} e^{i \frac{2\pi n \tau} \beta } J_{\ell+\frac 1 2} (\omega _{\ell j} r) Y_{\ell m} (\theta, \varphi ), \nn\eeq
with the spherical surface harmonics $Y_{\ell m} (\theta, \varphi)$ \cite{erde55b} solving $$-\frac 1 {\sin ^2 \theta} \frac{\partial^2} {\partial \varphi^2} - \frac 1 {\sin \theta } \frac \partial {\partial \theta} \sin \theta \frac \partial {\partial \theta} Y_{\ell m} (\theta, \varphi) = \ell (\ell +1) Y_{\ell m} (\theta , \varphi ),$$ and with the Bessel function $J_\nu (z)$, which is the regular solution of the differential equation \cite{grad65b}
$$\frac{d^2 J_\nu (z)} {dz^2} + \frac 1 z \frac{dJ_\nu (z)} {dz} + \left( 1 - \frac{\nu^2} {z^2} \right) J_\nu (z) =0.$$ Imposing the boundary condition on the unit sphere, \beq J_{\ell +\frac 1 2} (\omega_{\ell j} ) =0,\label{res4a}\eeq determines the eigenvalues. Namely, \beq \lambda_{n\ell j} ^2 = \left( \frac{2\pi n} \beta\right)^2 + \omega_{\ell j} ^2 + m^2 , \quad \quad n\in\Z, \quad \ell \in \N_0, \quad j\in \N. \label{res4} \eeq
This leads to the analysis of the zeta function \beq \zeta_P (s) = \sum_{n=-\infty}^\infty \sum_{\ell =0}^\infty \sum_{j=1}^\infty (2\ell +1) \left( p_n^2 + \omega_{\ell j}^2 + m^2\right)^{-s} , \label{res5}\eeq where we have used the standard abbreviation $p_n=(2\pi n)/\beta$. The factor $(2\ell +1)$ represents the multiplicity of eigenvalues for angular momentum $\ell$.

The zeroes $\omega_{\ell j}$ of the Bessel functions $J_{\ell + \frac 1 2}(\omega_{\ell j})$ are not known in closed form and thus we represent the $j$-summation using contour integrals. Starting with equation (\ref{res4a}) and following the argumentation of the previous section, this gives the identity \beq \zeta_P (s) = \sum_{n=-\infty}^\infty \sum_{\ell =0}^\infty (2\ell +1)
 \int\limits_\gamma \frac{d\lambda}{2\pi i} (p_n^2+\lambda^2+m^2)^{-s} \frac d {d\lambda} \ln J_{\ell +\frac 1 2} (\lambda ) , \label{res5a}\eeq valid for $\Re s>2$. The contour $\gamma$ runs counterclockwise and must enclose all the solutions of (\ref{res4a}) on the positive real axis. The next step is to shift the contour and place it along the imaginary axis. As $\lambda \to 0$ we observe that to leading order $J_\nu (\lambda ) \sim \lambda^\nu / (2^\nu \Gamma (\nu +1))$ such that the integrand diverges in this limit. Therefore, we include an additional factor $k^{-\ell - 1/2}$ in the logarithm in order to avoid contributions coming from the origin. Because there is no additional pole enclosed, this does not change the result. Furthermore we should note that the integrand has branch cuts starting at $\lambda = \pm i (p_n^2+m^2)$. Leaving out the $n,\ell$ summations for the moment and considering the $\lambda$-integration alone, we then obtain $(\nu =\ell + \frac 1 2)$
\beq \zeta_{P,n\ell} (s) &:=& \int\limits_\gamma \frac{d\lambda}{2\pi i} (p_n^2+\lambda^2+m^2)^{-s} \frac d {d\lambda} \ln \left[ \lambda^{-\nu} J_\nu (\lambda ) \right] \nn\\
&=&\frac{\sin \pi s} \pi \int\limits_{\sqrt{p_n^2+m^2}}^\infty dk \,\, (k^2 - p_n^2-m^2)^{-s} \frac d {dk} \ln \left[ k^{-(\ell + \frac 1 2)} I_{\ell + \frac 1 2} (k)\right],\label{res6}\eeq
where $J_\nu (ik) = e^{i\pi \nu} J_\nu (-ik)$ and $I_\nu (k) = e^{-i\nu \pi /2} J_\nu (ik)$ has been used \cite{grad65b}.

The next step is to add and subtract the asymptotic behavior of the integrand in (\ref{res6}). The relevant uniform asymptotics, after substituting $k=\nu z$ in the integral, is the Debye expansion of the Bessel functions \cite{abra70b}.
We have
\beq
I_{\nu} (\nu z) \sim \frac 1 {\sqrt{2\pi \nu}}\frac{e^{\nu
\eta}}{(1+z^2)^{\frac 1 4}}\left[1+\sum_{k=1}^{\infty} \frac{u_k (t)}
{\nu ^k}\right],\label{res7}
\eeq
with $t=1/\sqrt{1+z^2}$ and $\eta =\sqrt{1+z^2}+\ln
[z/(1+\sqrt{1+z^2})]$.
The first few coefficients are listed in \cite{abra70b},
higher coefficients are immediately obtained by using the recursion
\cite{abra70b}
\beq
u_{k+1} (t) =\frac 1 2 t^2 (1-t^2) u'_k (t) +\frac 1 8 \int\limits_0^t
d\tau\,\, (1-5\tau^2 ) u_k (\tau ),\label{res8}
\eeq
starting with $u_0 (t) =1$. As is clear, all the $u_k (t)$ are
polynomials in $t$.
The same holds for
the coefficients $D_n (t)$ defined by
\beq
\ln \left[1+\sum_{k=1}^{\infty} \frac{u_k (t)}{\nu ^k}\right] \sim
\sum_{n=1}^{\infty} \frac{D_n (t)}{\nu ^n}  .\label{res9}
\eeq
The polynomials $u_k (t)$ as well as $D_n (t)$
are easily found with the help of a simple computer program.
As we will see below, we need the first three terms in the expansion (\ref{res9}). Explicitly we have
\beq
D_1 (t) &=& \frac 1 8 t -\frac 5 {24} t^3 , \nn\\
D_2 (t) &=& \frac 1 {16} t^2 -\frac 3 8 t^4 +\frac 5 {16} t^6,\label{res10}\\
D_3 (t) &=& \frac{25}{384} t^3 - \frac{531}{640} t^5 + \frac{221}{128} t^7 - \frac{1105}{1152} t^9.\nn
\eeq
Adding and subtracting these terms in (\ref{res6})
allows us to rewrite the zeta function as
$$ \zeta _P(s) = \zeta_{P,f} (s) + \zeta_{P,as} (s),$$ where
\beq \zeta_{P,f} (s) &=& \frac{\sin \pi s} \pi \sum_{n=-\infty} ^\infty \sum_{\ell =0} ^\infty (2\ell +1) \int\limits_{\sqrt{\frac{p_n^2+m^2} {\nu^2} }}^\infty dz \,\, (z^2\nu^2-p_n^2-m^2)^{-s} \nn\\
& &\hspace{-1.0cm}\times \frac d {dz} \left\{ \ln [ z^{-\nu} I_\nu (\nu z) ]
- \ln \left[ \frac{z^{-\nu}}{\sqrt {2\pi \nu}} \frac{e^{\nu \eta}}{(1+z^2)^{1/4}} \right] - \frac {D_1 (t)} \nu - \frac{D_2 (t)} {\nu^2} - \frac{D_3 (t)} { \nu^3} \right\}, \label{res10a}\\
\zeta_{P,as} (s) &=& \frac{\sin \pi s} \pi \sum_{n=-\infty} ^\infty \sum_{\ell =0} ^\infty (2\ell +1) \int\limits_{\sqrt{\frac{p_n^2+m^2} {\nu^2} }}^\infty dz \,\, (z^2\nu^2-p_n^2-m^2)^{-s} \nn\\
& &\times\frac d {dz} \left\{  \ln \left[ \frac{z^{-\nu}}{\sqrt {2\pi \nu}} \frac{e^{\nu \eta}}{(1+z^2)^{1/4}} \right] + \frac {D_1 (t)} \nu + \frac{D_2 (t)} {\nu^2} + \frac{D_3 (t)} { \nu^3} \right\}.\label{res10b}\eeq
The number of terms subtracted in (\ref{res10a}) is chosen such that $\zeta_{P,f} (s)$ is analytic about $s=0$. The contributions from the asymptotics collected in (\ref{res10b}) are simple enough for an analytical continuation to be found.
Although it would be possible to proceed just with the contribution from inside the ball, in order to make the calculation as transparent and unambiguous as possible (as far as the interpretation of results goes) let us add the contribution from outside the ball.

The exterior of the ball, once the free Minkowski space contribution is subtracted, yields the starting point (\ref{res6}) with the replacement $k^{-\nu}I_\nu \to k^\nu K_\nu$ \cite{bord97-56-4896}. In this case the relevant uniform asymptotics is \cite{abra70b}
\beq
K_{\nu} (\nu z) \sim \sqrt {\frac \pi {2 \nu}} \frac{e^{-\nu
\eta}}{(1+z^2)^{\frac 1 4}}\left[1+\sum_{k=1}^{\infty} (-1)^k \frac{u_k (t)}
{\nu ^k}\right],\label{res11}
\eeq
where the notation is as in (\ref{res7}).
This produces the analogous splitting of the zeta function for the exterior space. Due to the characteristic sign changes in the asymptotics of $I_\nu$ and $K_\nu$, adding up the interior and exterior contributions several cancelations take place. As a result, the zeta function for the total space has the form
$$\zeta_{tot} (s) = \zeta_{tot,f} (s) + \zeta_{tot,as} (s)$$ with
\beq \zeta_{tot,f} (s) &=& \frac{\sin \pi s} \pi \sum_{n=-\infty} ^\infty \sum_{\ell =0} ^\infty (2\ell +1) \int\limits_{\sqrt{\frac{p_n^2+m^2} {\nu^2} }}^\infty dz \,\, (z^2\nu^2-p_n^2-m^2)^{-s} \nn\\
& &\frac d {dz} \left\{ \ln [ I_\nu (\nu z) K_\nu (\nu z) ] + \ln (2 \nu) +\frac 1 2 \ln (1+z^2) - \frac 2 {\nu^2} D_2 (t) \right\}, \label{res11a}\\
\zeta_{tot,as} (s) &=& \frac{\sin \pi s} \pi \sum_{n=-\infty} ^\infty \sum_{\ell =0} ^\infty (2\ell +1) \int\limits_{\sqrt{\frac{p_n^2+m^2} {\nu^2} }}^\infty dz \,\, (z^2\nu^2-p_n^2-m^2)^{-s} \nn\\
& &\frac d {dz} \left\{  - \ln (2 \nu) -\frac 1 2 \ln (1+z^2) + \frac 2 {\nu^2} D_2 (t) \right\}. \label{res11b}\eeq
By construction, $\zeta_{tot,f} (s)$ is analytic about $s=0$ and one immediately finds \beq \zeta_{tot,f} ' (0) &=& - \snmp \slnp (2\ell +1)\nn\\
 & &\left. \hspace{-.5cm}\times \left[ \ln \left( I_\nu (\nu z) K_\nu (\nu z) \right)+ \ln (2\nu) + \frac 1 2 \ln\left( 1+z^2\right) - \frac 2 {\nu^2} D_2 (t) \right] \right|_{z=\sqrt{\frac{p_n^2+m^2} {\nu^2}}},\label{res12}\eeq
with $t=1/\sqrt{1+z^2}$ as defined earlier. Although one could use (\ref{res12}) for numerical evaluation, further simplifications are possible.
Following \cite{dunn09-42-075402} we rewrite this expression according to \beq 1+z^2 = 1+ \frac{p_n^2 +m^2} {\nu^2} = \left( 1 + \frac{p_n^2}{\nu^2}\right) \left( 1+ \frac{m^2}{\nu^2+p_n^2}\right) . \label{res12a}\eeq
The advantage of the right hand side is that it can be expanded further for $\nu^2\to\infty$ or $p_n^2 \to \infty$ or both. This will allow us to subtract exactly the behavior that makes the double series convergent; the oversubtraction immanent in (\ref{res12}) can then be avoided. It is expected that
expanding the rightmost factor further for $\nu^2+p_n^2\gg 1$ leads to considerable cancelations when combined with $\zeta_{tot,as} ' (0)$ \cite{dunn09-42-075402}.

We split the asymptotic terms in (\ref{res12}) into those strictly needed to make the sums convergent
and those that ultimately will not contribute.
For example, we expand according to \beq
& &\left.\ln (1+z^2)\right|_{\sqrt{\frac{p_n^2+m^2}{\nu^2}}}=\ln \left( 1+ \frac{p_n^2+m^2}{\nu^2}\right) \nn\\
&=& \ln \left( 1 + \frac{p_n^2}{\nu^2}\right) + \ln \left( 1 + \frac{m^2}{\nu^2+p_n^2}\right) \nn\\
&=& \ln \left( 1 + \frac{p_n^2}{\nu^2}\right)+ \frac{m^2}{\nu^2+p_n^2} + \left[ \ln \left( 1+\frac{m^2}{\nu^2+p_n^2}\right) - \frac{m^2}{\nu^2+p_n^2}\right] .\nn\eeq
The first two terms have to be subtracted in (\ref{res12}) in order to make the summations convergent. The terms in square brackets are of the order ${\mathcal O} (1/(\nu^2+p_n^2)^2)$ and even after performing the summations in (\ref{res12}) a finite result follows. Thus the first two terms represent a minimal set of terms to be subtracted in (\ref{res12}) in order to make the sums finite. This minimal set of necessary terms will be called $\ln f_\ell ^{asym,(1)} (i\sqrt{p_n^2+m^2})$. The last two terms can be summed separately yielding a finite answer; those terms are summarized under $\ln f_\ell ^{asym,(2)} (i\sqrt{p_n^2+m^2})$. One can proceed along the same lines for all other terms.
With the definition \beq \hspace{1cm}\ln f_\ell^{asym} (i\sqrt{p_n^2+m^2}) &=& \left.- \ln (2\nu ) - \frac 1 2 \ln (1+z^2)+ \frac 2 {\nu^2} D_2 (t) \right|_{z=\sqrt{ \frac{p_n^2+m^2} {\nu^2} }} \nn\\
& &\hspace{-1.0cm}= \ln f_\ell^{asym,(1)} (i\sqrt{p_n^2+m^2})+ \ln f_\ell^{asym,(2)} (i\sqrt{p_n^2+m^2}) \label{res13}\eeq
the splitting is
\beq \ln f_\ell^{asym, (1)} (i \sqrt {p_n^2+m^2} ) &=& - \ln (2\nu ) - \frac 1 2 \ln \pnnu - \frac 1 2 \mnup \nn\\
& &+ \frac 2 {\nu^2} \left[ \frac 1 {16} \frac 1 {\pnnu} - \frac 3 8 \frac 1 {\pnnu ^2} + \frac 5 {16} \frac 1 {\pnnu ^3} \right] , \label{res14}\\
\ln f_\ell ^{asym, (2)} (i\sqrt {p_n^2+m^2} ) &=& - \frac 1 2 \ln \left( 1 + \mnup \right) + \frac 1 2 \mnup \nn\\
& &+ \frac 2 {\nu^2} \left\{ \frac 1 {16} \frac 1 {\pnnu} \left[ \frac 1 {1+\mnup} -1\right]\right.\nn\\
& &- \frac 3 8 \frac 1 {\pnnu^2} \left[ \frac 1 {\left(1+\mnup\right)^2} -1\right]\label{res15}\\
& &\left.+ \frac 5 {16} \frac 1 {\pnnu^3} \left[ \frac 1 {\left(1+\mnup\right)^3} -1\right]\right\}.\nn\eeq
We have used the given notation for the asymptotics to make a comparison with \cite{dunn09-42-075402} as easy as possible.
With these asymptotic quantities we rewrite $\zeta _{tot,f} ' (0)$ as
\beq \zeta _ {tot,f} ' (0) &=& - \snmp \slnp (2 \ell +1) \nn\\
& &\times \left[ \ln (I_\nu (\sqrt{ p_n^2 + m^2}) K_\nu (\sqrt{p_n^2+m^2}))- \ln f_\ell ^{asym , (1)} (i \sqrt{ p_n^2+m^2} ) \right]\nn\\
& &+ \snmp \slnp (2 \ell +1)\ln f_\ell ^{asym , (2)} (i \sqrt{ p_n^2+m^2} ) .\label{res16}\eeq
Let us next analyze $\zeta_{tot,as} ' (0)$. To analyze $\zeta_{tot,as} (s)$, equation (\ref{res11b}), further we use the integrals
\beq \int\limits_{\sqrt{p_n^2+m^2}}^\infty du \,\, (u^2-p_n^2-m^2)^{-s} \frac d {du} \ln \left( 1 + \frac{u^2} {\nu^2} \right)& =&\frac \pi {\sin \pi s} (m^2+\nu^2+p_n^2)^{-s} , \nn\\
 \int\limits_{\sqrt{p_n^2+m^2}}^\infty du \,\, (u^2-p_n^2-m^2)^{-s} \frac d {du} \left( 1 + \frac{u^2} {\nu^2} \right)^{-N/2} &=&\nn\\
  & &\hspace{-2.0cm} - \frac{\pi \Gamma \left( s + \frac N 2 \right)}{\sin (\pi s) \Gamma \left( \frac N 2 \right) \Gamma (s) } \frac{\nu^{-2s}}{\left( 1 + \pmnu\right)^{s+\frac N 2 }},\nn\eeq which are the relevant ones after substituting $z\nu =u$. This shows
\beq \zeta _{tot,as} (s) &=& - \sln \nu^{1-2s} \left( 1 + \pmnu \right)^{-s}\nn\\
& &- \frac s 4 \sln \nu^{-2s-1} \left( 1+ \pmnu \right)^{-s-1} \nn\\
& &+ \frac {3s (s+1)} 2 \sln \nu^{-2s-1} \left( 1+ \pmnu \right)^{-s-2} \label{res17}\\
& &- \frac{5s (s+1) (s+2)} 8 \sln \nu^{-2s-1} \left( 1 + \pmnu \right)^{-s-3} .\nn\eeq
To each of these terms we apply the rewriting (\ref{res12a}). Intermediate expressions are relatively lengthy and we explain details only for the first term. We proceed as for the splitting in (\ref{res14}) and (\ref{res15}) and write
\beq & &- \sln \nu^{1-2s} \left(1+\pmnu \right)^{-s}  \nn\\
&=&- \sln \nu^{1-2s} \pnnu ^{-s} \left(1+\mnup \right)^{-s}\nn\\
&=& - \sln \frac \nu {(\nu^2+p_n^2)^s} \left[ \left( 1+\mnup \right)^{-s} -1+s \mnup + 1-s\mnup\right] \nn\\
&=& - \sln \frac \nu {(\nu^2+p_n^2)^s} \left[ \left( 1+\mnup \right)^{-s} - 1 + s \mnup \right] \nn\\
& &-\frac 1 2 \sln \frac{2\nu} {(\nu^2+p_n^2)^s} + \frac {sm^2} 2 \sln \frac{2\nu} {(\nu^2+p_n^2)^{s+1}}.\label{res17a}\eeq
Note, that the first line is seen to be analytic about $s=0$. We have subtracted the minimal number of terms to make the sums convergent. The remaining terms represent Epstein type zeta functions,
\beq E^{(k)} (s,a) =  \sln 2\nu \frac{\nu^k}{(\nu^2+a^2n^2)^s}, \label{res18}\eeq
the analytical continuation of which is well understood. Performing a Poisson resummation in the $n$-summation \cite{ambj83-147-1,eliz90-31-170,kirs93-26-2421} yields
\beq E^{(k)} (s,a) &=& \frac{2 \sqrt \pi} {a} \frac{\Gamma \left( s-\frac 1 2\right)}{\Gamma (s)} \zeta_H (2s-k-2, 1/2) \nn\\
& &+ \frac{8 \pi^s} { \Gamma (s) a^{s+1/2}} \sum_{\ell =0}^\infty \nu^{k+\frac 3 2 -s} \sum_{n=1}^\infty n^{s-\frac 1 2} K_{\frac 1 2 -s} \left( 2\pi \nu \frac n a\right) .\label{res19} \eeq
The first line has poles at $s=1/2-j$, $j\in\N_0$, and for $s=(k+3)/2$, the second line is analytic for $s\in\C$.

In terms of these Epstein functions, in equation (\ref{res17a}) we have shown \beq - \sln \nu^{1-2s} \left( 1 + \pmnu \right)^{-s} &=& \nn\\
& &\hspace{-3.8cm}-\frac 1 2 E^{(0)} \left( s,\frac{2\pi} \beta\right) + \frac{sm^2} 2 E^{(0)} \left( s+1, \frac{2\pi} \beta\right)\nn\\
& &\hspace{-3.8cm}- \sln \frac \nu {(\nu^2+p_n^2)^s} \left[ \left( 1+\mnup\right)^{-s} -1+s \mnup\right] .\label{res20}\eeq
Noting from equation (\ref{res19}) that $E^{(0)} (s,a)$ and $E^{(0)} (s+1,a)$ are analytic about $s=0$, we get \beq - \frac d {ds}  \sln \nu^{1-2s} \left( 1 + \pmnu \right)^{-s}
&=& - \frac 1 2 E^{(0)\prime} \left( 0, \frac{2\pi} \beta\right) + \frac 1 2 m^2 E^{(0)\prime} \left(1,\frac{2\pi} \beta \right) \nn\\
& &\hspace{-3cm}-\sln \nu \left[ - \ln \left( 1+\mnup \right) + \mnup \right].\label{res21}\eeq
The last term on the right cancels the first line from $\ln f_\ell ^{asym,(2)} (i\sqrt{p_n^2+m^2})$ in equation (\ref{res15}), the remaining terms are easily found from (\ref{res19}).

One can proceed in exactly this way for the other terms in $\zeta_{tot,as}(s)$; there are always terms that cancel with terms from $\ln f_\ell ^{asym,(2)} (i\sqrt{p_n^2+m^2})$ in equation (\ref{res15}) and terms expressible using the Epstein type zeta functions given in equation (\ref{res18}). Adding up all contributions, the second line in equation (\ref{res16}) completely cancels and we obtain the following closed form for the finite-$T$ zeta function,
\beq \zeta_{tot} ' (0) &=& - \sln 2\nu \left[ \ln I_\nu (\sqrt{p_n^2+m^2}) K_\nu (\sqrt{p_n^2+m^2})\right.\nn\\
& & + \ln (\nu^2+p_n^2) + \mnup \nn\\
& &\left. - \frac 1 8 \frac 1 {\nu^2+p_n^2} \left( 1 - \frac{6\nu^2} {(\nu^2+p_n^2)^2} + \frac{5\nu^4}{(\nu^2+p_n^2)^3}\right)\right]\nn\\
& &-\frac 1 2 E^{(0)\prime} \left( 0, \frac{2\pi} \beta \right) + \frac 1 2 m^2 E^{(0)} \left( 1 , \frac{2\pi} \beta \right) \nn\\
& &+ \frac 3 4 E^{(2)} \left( 2, \frac{2\pi} \beta \right) - \frac 5 8 E^{(4)} \left( 3, \frac{2\pi} \beta \right) .\label{res22}\eeq
From equation (\ref{res19}) it is clear that the Epstein type zeta functions contain zero temperature contributions to the Casimir energy (first line in (\ref{res19})) and
exponentially damped contributions for small temperature described by the Bessel functions (second line in (\ref{res19})). As it turns out, the zero temperature contributions from the Epstein type zeta functions in (\ref{res22}) all vanish. The remaining zero temperature contributions in (\ref{res22}) are found replacing the Riemann sum over $n$ by an integral,
$$\sum_{n=-\infty}^\infty f(n) \Longrightarrow \frac \beta {2\pi} \int\limits_{-\infty}^\infty dp f(p).$$ As $\beta \to 0$ this shows
\beq \frac 1 \beta \zeta_{tot} ' (0) &=& - \frac 1 {2\pi} \int\limits_{-\infty}^\infty dp \sum_{\ell =0}^\infty 2 \nu \left[ \ln I_\nu ( \sqrt{p^2+m^2}) K_\nu (\sqrt{p^2+m^2}) \right.\nn\\
& +&\left. \ln (\nu^2+p^2) + \frac{m^2}{\nu^2+p^2}
- \frac 1 8 \frac 1 {\nu^2+p^2} \left( 1- \frac{6\nu^2}{(\nu^2+p^2)^2} + \frac{5\nu^4}{\nu^2+p^2)^3}\right)\right],\label{res23}\eeq
from which the Casimir energy (\ref{res3}) is trivially obtained.
The result is much simpler than previous results given \cite{defr94-50-2908,bord97-56-4896} and a numerical evaluation could easily be performed.
\section{Functional determinant on a two dimensional torus}
As our next example let us consider a two dimensional torus $S^1\times S^1$. For convenience we choose the perimeter of the circles to be $1$. The relevant eigenvalue problem to be considered then is $$ P \phi_\lambda (x,y) := \left( - \frac{\partial ^2}{\partial x^2} - \frac{\partial ^2} {\partial y^2} \right) \phi_\lambda (x,y) = \lambda^2 \phi_\lambda (x,y) , $$ and we choose periodic boundary conditions \beq \phi_\lambda (x,y ) = \phi (x+1, y ) ,  & & \quad \phi_\lambda (x,y) = \phi _\lambda (x,y+1) , \nn\\
\frac {\partial \phi_\lambda (x,y) } {\partial x} = \frac{\partial \phi (x+1,y)} {\partial x} , & & \quad
\frac {\partial \phi_\lambda (x,y) } {\partial y} = \frac{\partial \phi (x,y+1)} {\partial y}. \nn\eeq
Eigenfunctions and eigenvalues clearly are \beq \phi_{m,n} (x,y) = e^{-2\pi i mx} \cdot e^{-2\pi iny} , \quad \quad \lambda^2 = (2\pi )^2 (m^2+n^2), \quad n,m\in\Z.\nn\eeq
The related zeta function then reads
\beq \zeta_P (s) = (2\pi)^{-2s}\sum_{(m,n)\in\Z ^2/\{(0,0)\}} (m^2+n^2)^{-s}; \label{res24}\eeq
note that the zero mode $m=n=0$ has to be omitted in the summation to make $\zeta _P(s)$ well defined. The zeta function in equation (\ref{res24}) is an Epstein zeta function and $\zeta _P' (0)$ can be evaluated using the Kronecker limit formula \cite{epst03-56-615,epst07-63-205}. Here, we apply the contour approach previously outlined which simplifies the calculation.

Instead of using the fact that the eigenvalues can be given in closed form, we proceed differently. We say that $$\lambda^2 = (2\pi)^2 (n^2+k^2),\quad \quad n\in\Z,$$ where $k$ is determined as a solution to the equation \beq e^{\pi i k} - e^{-\pi ik} =0.\label{res24a}\eeq
Of course, solutions are given by $k\in\Z$ and the correct eigenvalues follow. Using equation (\ref{res24a}) determining the eigenvalues in the way we have used equations (\ref{res02}) and (\ref{res4a}), the zeta function can be represented as the contour integral \beq \zeta_P (s) &=& 4 \sum_{n=1}^\infty \int\limits_\gamma \frac{dk}{2\pi i} (2\pi)^{-2s} (n^2+k^2)^{-s} \frac d {dk} \ln \left( \frac{e^{\pi ik} - e^{-\pi i k} } {2\pi ik} \right) \nn\\
& &+ 4 (2\pi)^{-2s}\zeta_R (2s).\label{res25}\eeq
The second line represents the part where one of the two indices $m$ or $n$ is zero in equation (\ref{res24}). The first line represents the remaining contributions. The factor of $4$ is a result of summing over positive $n$ only and because the contour $\gamma$ is supposed to enclose positive integers only.
The reason that we have used $$\frac{e^{\pi ik} - e^{-\pi ik}} {2\pi ik} $$ instead of the above equation (\ref{res24a}) is that $$\lim _{k\to 0} \frac{e^{\pi ik} - e^{-\pi ik}} {2\pi ik} =1,$$ which will allow us to shift the contour in a way as to include the origin; see the discussion below equation (\ref{res5a}). Let us evaluate the contour integral $$\zeta_n (s) = \int\limits_\gamma \frac{dk}{2\pi i} (2\pi)^{-2s} (n^2+k^2)^{-s} \frac d {dk} \ln \left( \frac{e^{\pi ik} - e^{-\pi i k} } {2\pi ik} \right).$$ Substituting $k=\sqrt z$ and deforming the contour to the negative real axis along the lines described previously, an intermediate result is \beq \zeta _n (s) =  (2\pi )^{-2s} \frac{\sin \pi s} \pi \int\limits_{n^2}^\infty dz (z-n^2)^{-s} \frac d {dz} \ln \left( \frac{e^{\pi \sqrt z} - e^{-\pi \sqrt z}}{2\pi \sqrt z} \right) .\label{res26}\eeq
From the behavior of the integrand as $z\to \infty$ and $z\to n^2$ this representation is seen to be valid for $1/2 < \Re s < 1$. In order to construct the analytical continuation to a neighborhood of $s=0$ we note that $$\frac{e^{\pi \sqrt z} - e^{-\pi \sqrt z} } {2\pi \sqrt z} = \frac{e^{\pi \sqrt z}} {2\pi \sqrt z} \left( 1-e^{-2\pi \sqrt z}\right).$$ We therefore write \beq \zeta_n (s) &=&  (2\pi )^{-2s} \frac{\sin \pi s} \pi \int\limits_{n^2}^\infty dz (z-n^2)^{-s} \frac d {dz} \ln \left( \frac{e^{\pi \sqrt z}}{2\pi \sqrt z}\right) \nn\\
& & + (2\pi )^{-2s} \frac{\sin \pi s} \pi \int\limits_{n^2}^\infty dz (z-n^2)^{-s} \frac d {dz} \ln \left(1-e^{-2\pi \sqrt z}\right). \nn\eeq
The first line is evaluated using \beq \int\limits_{n^2}^\infty dz \frac{(z-n^2)^{-s}}{\sqrt z} &=&\frac{n^{1-2s}}{\sqrt \pi} \Gamma (1-s) \Gamma \left( - \frac 1 2 +s \right) , \nn\\
\int\limits_{n^2}^\infty dz \frac{(z-n^2)^{-s}}{ z} &=& \frac{\pi n^{-2s}}{\sin \pi s} .\nn\eeq
With the identity \cite{grad65b}
$$\frac{\sin\pi s} \pi \,\,\Gamma (1-s) = \frac 1 {\Gamma (s)} $$ this produces
\beq \zeta_n(s) &=& \frac 1 4 
(2\pi)^{-2s+1} \frac{n^{1-2s} \Gamma \left( - \frac 1 2 +s \right)}{\sqrt \pi \Gamma (s)} - \frac 1 2 (2\pi)^{-2s} n^{-2s} \nn\\
& & +(2\pi)^{-2s} \frac{\sin \pi s} \pi \int\limits_{n^2}^\infty dz (z-n^2)^{-s} \frac d {dz} \ln \left( 1-e^{-2\pi \sqrt z} \right).\nn\eeq
This is the form that allows the sum over $n$ to be (partly) performed and it shows
\beq \zeta _P(s) &=& 4 (2\pi)^{-2s} \frac{\sin \pi s } \pi \sum_{n=1}^\infty \,\,\,\int\limits_{n^2}^\infty dz (z-n^2)^{-s} \frac d {dz} \ln \left( 1-e^{-2\pi \sqrt z}\right) \nn\\
& & -2 (2\pi)^{-2s} \zeta _R (2s) + (2\pi)^{-2s+1} \frac{\Gamma \left( - \frac 1 2 +s \right)}{\sqrt \pi \Gamma (s)} \zeta_R (2s-1) .\label{rees27}\eeq
This form allows for the evaluation of $\zeta_P ' (0)$. From known elementary properties of the $\Gamma$-function and the zeta function of Riemann \cite{grad65b} we obtain
\beq \left.\frac d {ds} \right|_{s=0} \left[ (2\pi)^{-2s} \zeta _R (2s)\right] &=& - 2 \ln (2\pi) \zeta _R (0) + 2 \zeta_R ' (0) =0,\nn\\
\left.\frac d {ds} \right|_{s=0} \left[ (2\pi)^{-2s+1} \frac{\Gamma \left( - \frac 1 2 +s \right)}{\sqrt \pi \Gamma (s)} \zeta_R (2s-1)\right]& =& \frac \pi 3.\nn\eeq
The first line in (\ref{rees27}) is also easily evaluated because \beq & &\left.\frac d {ds}\right|_{s=0}
4 (2\pi)^{-2s} \frac{\sin \pi s } \pi \sum_{n=1}^\infty \,\,\int\limits_{n^2}^\infty dz (z-n^2)^{-s} \frac d {dz} \ln \left( 1-e^{-2\pi \sqrt z}\right) =\nn\\
& & 4\sum_{n=1}^\infty \,\,\int\limits_{n^2}^\infty dz \frac d {dz} \ln \left( 1-e^{-2\pi \sqrt z}\right)
=- 4 \sum_{n=1}^\infty \ln \left( 1-e^{-2\pi n}\right).\nn\eeq
This can be reexpressed using the Dedekind eta function
$$\eta (\tau ) := e^{i\pi\tau /12} \prod_{n=1}^\infty \left( 1-e^{2\pi in\tau} \right) $$ for $\tau \in \C$, $\Re \tau >0$. The relation relevant for us follows by setting $\tau =i$, $$\ln | \eta (i) | ^4 = - \frac \pi 3 + 4 \sum_{n=1}^\infty \ln \left( 1- e^{-2\pi n}\right).$$ Adding up all contributions for $\zeta_P ' (0)$, the final answer reads \beq\zeta _P' (0) = \frac \pi 3 - 4 \sum_{n=1}^\infty \ln \left( 1-e^{-2\pi n}\right) = - \ln | \eta (i)|^4\label{res28}\eeq
in agreement with known answers; see, e.g., \cite{polc98b,will03b}.
\section{Conclusions}
In this article we have shown that contour integrals are very useful and effective tools for the evaluation of determinants of differential operators. Although the results look very simple only in one dimension, see equation (\ref{res08a}), for particular configurations also in higher dimensions closed answers can be found suitable for numerical evaluation, see equations (\ref{res23}) and (\ref{res28}). Here we have provided answers only for the torus and a spherically symmetric situation. But the same ideas should apply when separability of the partial differential equations in other coordinate systems is possible. Results in this direction will be presented elsewhere.

\section*{Acknowledgement}
KK acknowledges support by the NSF through grant PHY-0757791. Part of the work was done while the author
enjoyed the hospitality and partial support of the
Department of Physics and Astronomy of
the University of Oklahoma. Thanks go in particular to Kimball Milton and his group who
made this very pleasant and exciting visit possible.



\end{document}